\documentclass[11pt,twoside]{article}
\usepackage{asp2010}

\resetcounters

\begin{document}

\title{A Relic Star Cluster in the Sextans Dwarf Spheroidal Galaxy -- Implications for Early Star and Galaxy Formation}
\author{Torgny Karlsson$^{1,2}$ and Joss  Bland-Hawthorn$^2$
\affil{$^1$Department of Physics and Astronomy, Uppsala University, Box 516, 751 20 Uppsala, Sweden}
\affil{$^2$Sydney Institute for Astronomy, School of Physics, University of Sydney, NSW 2006, Australia}}

\begin{abstract}
We present tentative evidence for the existence of a dissolved star cluster in the Sextans dwarf spheroidal galaxy. In a sample of six stars, we identify three (possibly four) stars around  $[\mathrm{Fe}/\mathrm{H}] = -2.7$ that are highly clustered in a multi-dimensional chemical abundance space. The estimated initial stellar mass of the cluster is $M_{*,\mathrm{init}} = 1.9^{+1.5}_{-0.9}~(1.6^{+1.2}_{-0.8})\times 10^5~\mathcal{M_{\odot}}$ assuming a Salpeter (Kroupa) initial mass function (IMF).  If corroborated by follow-up spectroscopy, this ancient star cluster at $[\mathrm{Fe}/\mathrm{H}] = -2.7$ is the most metal-poor system identified to date. Inspired by this finding, we also present a new way to interpret the cumulative metallicity functions of dwarf galaxies. From available observational data, we speculate that the ultra-faint dwarf galaxy population, or a significant fraction thereof, and the more luminous, classical dwarf spheroidal population were formed in different environments and would thus be distinct in origin.
\end{abstract}

\section{Introduction}\label{sect:intro}
Little is known about the star formation process at the earliest cosmic times. Recent hydrodynamics simulations suggest that primordial stars might have been formed in binary or small multiple systems (Stacy, Greif, \& Bromm 2010\nocite{stac10}). Assuming turbulent initial conditions, Clark et al. (2011\nocite{clar11}) showed that primordial stars could even have been formed in small, dense clusters. In their simulations, Wise et al. (2012\nocite{wise12}) found that most Pop II clusters below $[\mathrm{Z}/\mathrm{H}] = -2$ were formed with masses in the range $5 \times10^2 \lesssim M_*/\mathcal{M_{\odot}} \lesssim 10^4$, with a relative frequency resembling that of what is found today in the Galactic Disk (see their Fig. 6).

The most metal-poor star clusters known to date have an iron abundance just below $[\mathrm{Fe}/\mathrm{H}]=-2$. One of the globular clusters (Cluster 1) in the Fornax dwarf spheroidal (dSph) currently holds the record with a metallicity $[\mathrm{Fe}/\mathrm{H}]=-2.5$ (Letarte et al. 2006\nocite{leta06}). However, not all clusters have survived as gravitationally bound objects to the present epoch.  Much like the Galactic Halo, dwarf galaxies have stars with metallicities well below $[\mathrm{Fe}/\mathrm{H}]=-3$ (Kirby et al. 2008\nocite{kirb08}; Starkenburg et al. 2010\nocite{star10}). The relatively simple environments of the low mass dwarf galaxies raises the prospect of identifying disrupted star clusters at much lower metallicity than has been possible before through the technique of chemical tagging (Bland-Hawthorn et al. 2010a,b\nocite{blan10a,blan10b}). In the extension, this also gives us a unique tool to probe the formation and early evolution of the present dwarf galaxy population.

In the following sections, we will demonstrate the possible existence, estimate the initial mass, and discuss the likely nature of a disrupted stellar cluster at \mbox{$[\mathrm{Fe}/\mathrm{H}]=-2.7$} in the Sextans dSph. We end by briefly explore the implications for near-field cosmology and the possibility that the population of ultra-faint dwarfs and the more luminous, classical dSphs have a different origin.

\begin{figure*}[t]
\resizebox{\hsize}{!}{\includegraphics[trim=11.6mm 12mm 34.2mm 32mm,clip]{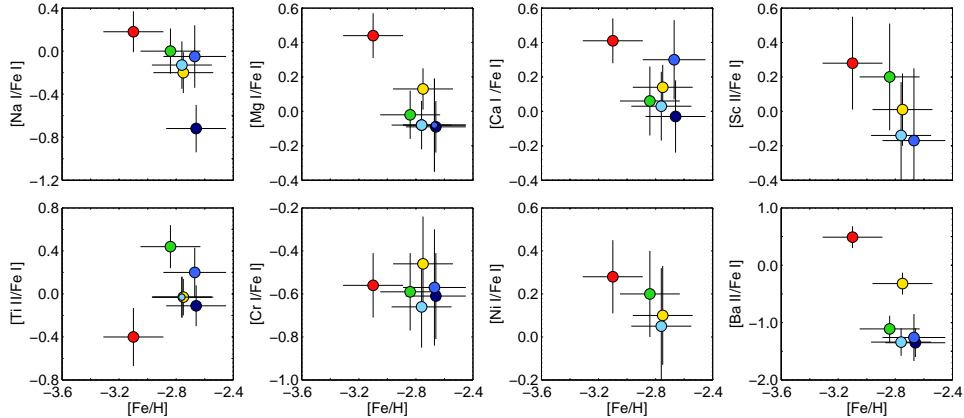}}
\caption{Chemical abundance ratios of six very metal-poor stars in the Sextans dSph. Abundance data are taken from Aoki et al. (2009). Stars are color coded according to their location in the [Mg/Fe] -- [Fe/H] diagram. The three stars with nearly identical [Mg/Fe] (colored in different shades of blue) also have very similar [Fe/H]. These stars all clump together in Ti, Cr, and Ba, as well.}
\label{fig:abund}
\nocite{aoki09}
\end{figure*}

\section{Observational evidences}\label{sect:obsev}
Based on high-resolution ($\mathcal{R} \simeq 40, 000$) spectroscopy, Aoki et al. (2009\nocite{aoki09}) recently determined the chemical abundances of six very metal-poor stars ($[\mathrm{Fe}/\mathrm{H}] < -2.5$) in the Sextans dSph. Their excellent analysis revealed a group of four stars that display a subzero [Mg/Fe] ratio with a small scatter around the weighted mean $\langle[\mathrm{Mg}/\mathrm{Fe}]\rangle = -0.06$ (see Fig. \ref{fig:abund}). This is in contrast to the Mg-to-Fe enhancements commonly observed in Galactic Halo stars of similar metallicity (Cayrel et al. 2004\nocite{cayr04}). A closer inspection of Fig. \ref{fig:abund} shows that three (colored in shades of blue) of the four stars more or less tightly clump together in Ti, Cr and Ba as well. Two out of three ``blue'' stars also have similar Na, Ca and Sc abundances (Sc was not measured in the dark blue star S~$10-14$).  As Aoki et al. (2009\nocite{aoki09}) points out, the higher [Ca/Fe] ratio in S~$14-98$ (medium blue) could simply be due to a larger observational uncertainty in Ca for this star. In terms of the chemistry, the star S~$11-37$ (color coded green in Fig. \ref{fig:abund}) may be regarded as a borderline case. However, since it is slightly off ($1-2 \sigma$) in Sc, Ti, and Ba we will not regard this star as a member of the ``blue'' group. We will return to the low Na abundance of S~$10-14$ in Sect. \ref{sect:natur}. 

\begin{figure}[t]
\resizebox{\hsize}{!}{\includegraphics[width=20mm, trim=1mm 8mm 3mm 15.2mm,clip]{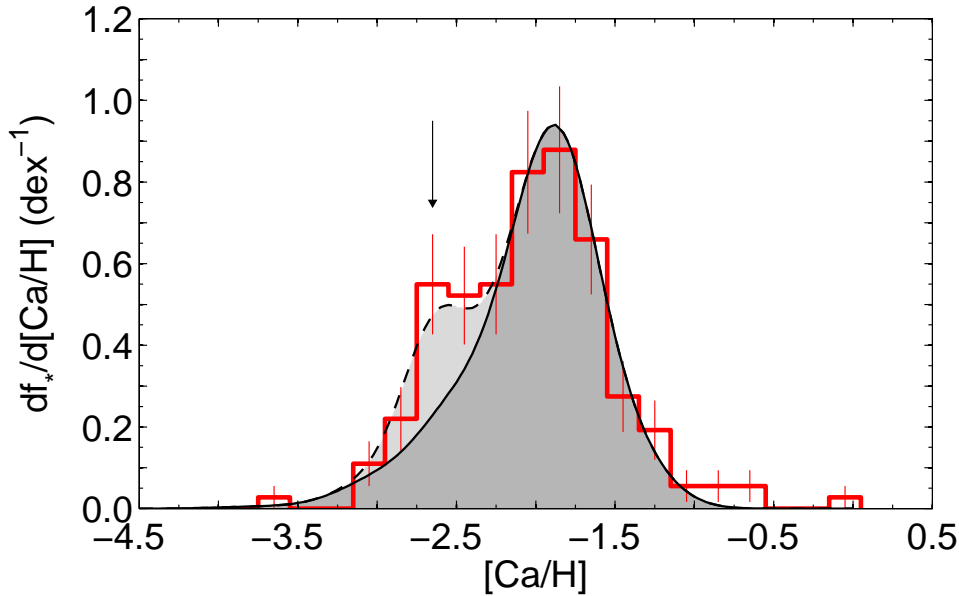}}
\caption{MDF of Sextans, as given by $\mathrm{d}f_{*}/\mathrm{d}[\mathrm{Ca}/\mathrm{H}]$. The quantity $f_{*}$ is the fraction of stars that fall below each [Ca/H] bin (1 dex). The black, solid line denotes the fiducial distribution of [Ca/H], predicted from our stochastic chemical evolution model while the black, dashed line denotes the corresponding distribution, including an $M_{\star}=2.0\times 10^5~\mathcal{M_{\odot}}$ star cluster at $[\mathrm{Ca}/\mathrm{H}] = -2.65$ (arrow). Both MDFs are convolved with a $\sigma = 0.2$ dex gaussian. The light-grey shaded area contains $12.3\%$ of the total stellar mass. The red step function (with Poissonian noise) shows the observed distribution of [Ca/H] (Starkenburg et al. 2010).}
\label{fig:mdf}
\nocite{star10}
\end{figure}

Starkenburg et al. (2010\nocite{star10}) determined re-calibrated metallicities for metal-poor stars in four classical dSphs, including Sextans. The new data reveal an excess of stars in the metallicity distribution function (MDF) of Sextans around $[\mathrm{Fe}/\mathrm{H}]\simeq -2.9$. Assuming a typical enhancement of $\langle[\mathrm{Ca}/\mathrm{Fe}]\rangle = +0.25$ in metal-poor stars (Starkenburg et al. 2010\nocite{star10}), this ``bump'' should be found at $[\mathrm{Ca}/\mathrm{H}] \simeq -2.65$ (see Fig. \ref{fig:mdf}). Interestingly, this is close to the weighted mean abundance of Ca for the three ``blue'' stars identified in the data of Aoki et al. (2009\nocite{aoki09}). 

If real, the ``bump'' could be due to the presence of a relic cluster. In conjunction with our finding above, we will argue that the ``blue'' stars in Fig. \ref{fig:abund} once were members of a chemically homogeneous star cluster that is now, at least partly, dissolved. This conclusion is also in accord with the detection of a kinematically cold substructure in the central region of Sextans, possibly originating from a remnant star cluster (Battaglia et al. 2011\nocite{batt11}). The substructure has an observed mean metallicity of $[\mathrm{Fe}/\mathrm{H}] \simeq -2.6$ (Battaglia et al. 2011\nocite{batt11}), which is very close to that of the ``blue'' group of stars.

\section{Stellar mass and nature of relic cluster}\label{sect:natur}
We first estimate the initial mass of the potential star cluster by measuring the size of the ``bump'' in the MDF of Sextans (Fig. \ref{fig:mdf}). By comparing the observed MDF with the MDF predicted from our stochastic chemical evolution model (for details, see Karlsson et al. 2012\nocite{karl12}), we determine the initial mass to $M_{*,\mathrm{init}}=2.0~(1.6) \times 10^5~\mathcal{M_{\odot}}$ for a Salpeter (Kroupa, Kroupa 2001\nocite{krou01}) initial mass function (IMF). At face value, the identification of the three ``blue'' stars in Fig. \ref{fig:abund}, if they do belong to a dispersed cluster, point towards a cluster mass in the range $2.2\times 10^5 \le M_{*,\mathrm{init}}/\mathcal{M_{\odot}}\le 3.6\times 10^5$. Given this conditional constraint, the combined estimate of the most probable mass of the dissolved cluster (with asymmetric uncertainties) is determined to $M_{*,\mathrm{init}} = 1.9^{+1.5}_{-0.9}~(1.6^{+1.2}_{-0.8})\times 10^5~\mathcal{M_{\odot}}$, for a Salpeter (Kroupa) IMF (Karlsson et al. 2012\nocite{karl12}).

Globular clusters show specific elemental abundance correlations, such as the Na-O and the Al-Mg anti-correlations. These are not present in open clusters and could therefore be used as discriminators between the two types of clusters. Interestingly,  the star S~$10-14$ (dark blue) has a significantly lower Na abundance than the other two potential cluster stars (see Fig. \ref{fig:abund}). This may be suggestive of an Na-O anti-correlation present in the data. The fact that one out of three stars show a low [Na/Fe] is also consistent with the fraction of globular cluster stars exhibiting primordial abundances (Carretta et al. 2009\nocite{carr09}). Along with the inferred initial mass, we conclude that the relic cluster might have been a globular cluster.  Further observations, including O and Al abundances are, however, required in order to readily determine the nature of the cluster.

\section{On the origin of ultra-faint dwarfs and dSphs}\label{sect:impl}
A comparison between models and available metallicity data for dSphs and ultra-faints suggests that the formation of clusters in the ultra-faints differed from that of the dSphs (Karlsson et al. 2012\nocite{karl12}). Together with a detected offset between the mean cumulative metallicity functions of the ultra-faints and the dSphs, and a putative offset in the $\langle[\mathrm{Mg}/\mathrm{Fe}]\rangle$ ratio, we speculate that these two types of galaxies probably were formed in different environments and would therefore be distinct in origin. A possible explanation to the apparent absence of clumping in the ultra-faints below $[\mathrm{Fe}/\mathrm{H}]= -2.5$ is that these galaxies were formed predominantly before the local Universe was reionized, while the dSphs formed mostly after reionization. However, whether there is a dichotomy, as suggested by the current observational data, or a continuous distribution of galaxies exhibiting a mix of properties should be investigated further.

\bibliography{refs}

\end{document}